# Versatile Quantum Machine Learning with an Ultra-low Power Photonic Quantum Reservoir Computer

Wei Wang[1,2*†], Zan Tang[1,2*], Menglong Fang[1,2], Daiqin Su[1,2†], Mile Gu[3], Jayne Thompson[4], Lip Ket Chin[1,2†], Hong Cai[1,2†], Leong-Chuan Kwek[5], Ai-Qun Liu[1,2]

[1] Research Institute for Quantum Technology (RIQT), The Hong Kong Polytechnic University, Hong Kong SAR, China

[2] Department of Electrical and Electronic Engineering, The Hong Kong Polytechnic University, Hong Kong SAR, China

[3] Nanyang Quantum Hub, School of Physical and Mathematical Sciences, Nanyang Technological University, Singapore

[4] College of Computing and Data Science, Nanyang Technological University, Singapore

[5] Centre for Quantum Technologies, National University of Singapore, Singapore

[*]Equivalent contribution

[†] Corresponding authors: wei-iqt.wang@polyu.edu.hk (W. W.); daiqin.su@polyu.edu.hk (D. Q. S.); helen.cai@polyu.edu.hk (H. C.); lkchin@polyu.edu.hk (L. K. C.)

## Abstract

Integrated photonic microprocessors provide high-bandwidth, massively parallel linear computation, but realizing nonlinear feature maps and temporal memory remain key challenges for machine learning. Conventional approaches rely on active tuning and additional nonlinear elements, increasing architectural complexity and power overhead. Here we demonstrate an integrated photonic quantum reservoir computer that achieves nonlinear mapping, fading memory, and task versatility without active tuning of the reservoir core. The same chip supports accurate static classification, dynamic prediction, and stable autonomous forecasting, establishing broad utility across both classification and temporal inference tasks. Competitive performance is retained in the zero-bias state, where all on-chip phase shifters are unpowered, eliminating active control and reducing computational power consumption to zero. This passive operation highlights a scalable route to multifunctional machine-learning hardware, where large-scale photonic quantum processors can be repurposed as reservoirs without reconfiguring their internal optical networks. By combining quantum-state encoding, multimode interferometric mixing, and photon-statistical readout, this architecture provides a physically grounded paradigm for low-power, large-scale quantum reservoir computing.

## Introduction

Quantum reservoir computing provides a hardware-efficient approach to quantum machine learning by exploiting the high-dimensional state space and intrinsic dynamics of quantum systems[1-3]. In contrast to fully parameterized quantum neural networks, quantum reservoir computing operates with a fixed quantum processor while training only a classical readout[4-6], thereby minimizing the need for repeated optimization and reconfiguration[7-9]. Demonstrations across nuclear-spin ensembles[10], correlated spin systems[11], superconducting quantum processors[12], and atom–cavity platforms[13], highlight the versatility of quantum dynamics as a computational resource. Realizing quantum reservoir computing as a general-purpose paradigm, however, requires scalable physical systems that generate nonlinear feature mappings, preserve temporal memory, and support diverse learning tasks with minimal hardware overhead.

Photonic systems offer a natural pathway to scalability, as quantum states of light provide high-dimensional information spaces and optical interference enables large-scale transformations with high bandwidth and parallelism[14-18]. To date, photonic reservoir computing has been demonstrated primarily in bulk-optical systems[19], with scaling to integrated architectures remaining a central challenge[18, 20-22]. Passive optical networks are intrinsically linear, whereas reservoir computing requires nonlinear transformations. Existing approaches, based on active components, nonlinear media, repeated optical-to-electrical conversion, or extensive input-dependent reconfiguration, introduce significant hardware and control complexity[20-25]. Moreover, it remains unclear whether a single integrated reservoir can serve as a reusable substrate for both static and dynamic machine-learning tasks.

Here, we develop and experimentally demonstrate the first integrated photonic quantum reservoir computer. Classical inputs are encoded in the squeezing parameters of vacuum states and then processed by a static linear optical circuit that serves as the reservoir. Photon-number statistics at the output provide high-dimensional features and effective nonlinear mappings, which are subsequently processed by a classical readout. Quantum dynamics generate a rich feature representation, while classical computation performs task-specific inference, without requiring re-optimization of the quantum hardware for each task. Using the same reservoir configuration, we demonstrate both static classification and dynamic time-series prediction, thereby demonstrating the versatility of the platform. Crucially, performance does not depend on a specially optimized optical network, thereby reducing energy overheads associated with active parameter control. These results establish an integrated photonic route toward a low-overhead, flexible, and energy-efficient quantum reservoir computer capable of addressing a broad range of machine learning problems, while also opening a pathway toward scalable architectures for large-scale quantum computing.

## Architecture of the quantum reservoir computer

The proposed quantum reservoir computer exploits the reservoir computing paradigm and supports a versatile range of machine-learning tasks, with the system-level architecture as conceptually depicted in **Fig. 1a**. For each machine-learning task, the data are stored in the random access memory (RAM), while the central processing unit (CPU) prepares the input data and converts them into the corresponding digital encoding parameters. These data are translated by the digital-to-analog converter into electrical control signals and loaded into the quantum microprocessor (QPU). Driven by a pump laser, the QPU maps classical inputs into quantum states, and the reservoir core performs computation. The reservoir output is detected and digitized before being returned to the CPU, where features are constructed and processed by a trained linear readout to generate the task output. For static tasks, the input consists of sample features or image pixels, and the output is a classification result. For dynamic tasks, the input is a time-series history window, and the output is a prediction of its future evolution. Task adaptation is implemented through input encoding and the trained linear readout, while the quantum reservoir core remains unchanged.

At the physical level, **Fig. 1b** illustrates the implementation of quantum reservoir computing within the photonic QPU. Task-dependent classical data, denoted as $\boldsymbol{x} = (x_1, x_2, \dots, x_M)^{\mathsf{T}}$ with the input dimension $M$, enter the encoder. Each component $x_i$ is mapped onto a squeezing parameter $r_i$ via $r_i = x_i$, thereby is encoded into a single-mode squeezed vacuum state. The overall input state to the reservoir core is expressed as

$$|\psi_{\text{in}}(\boldsymbol{x})\rangle = \otimes_{i=1}^{M} \hat{S}(r_i)|0\rangle, \quad (1)$$

where $|0\rangle$ is the vacuum state and the single-mode squeezing operator is expressed by $\hat{S}(r_i) = \exp\left[\frac{1}{2} r_i\left(\hat{a}_i^2 - \hat{a}_i^{\dagger 2}\right)\right]$, with $\hat{a}_i$ and $\hat{a}_i^\dagger$ the annihilation and creation operators of the $i$-th mode, respectively. The mean photon number of a single-mode squeezed vacuum state is $\sinh^2 r_i$, indicating that the input already enters observable statistics nonlinearly before reservoir processing.

The quantum reservoir core, implemented by a multimode linear-optical circuit, transforms the input independent single-mode squeezed vacuum states into an output multimode Gaussian state[26, 27]. This unitary evolution constitutes the reservoir dynamics in reservoir computing. In the Heisenberg picture, the transformation is described by $\hat{\boldsymbol{b}} = \boldsymbol{U}\hat{\boldsymbol{a}}$, where $\hat{\boldsymbol{a}} = (\hat{a}_1, \dots, \hat{a}_M)^\mathsf{T}$ and $\hat{\boldsymbol{b}} = (\hat{b}_1, \dots, \hat{b}_M)^\mathsf{T}$ are the input and output mode operators, respectively, and $\boldsymbol{U}$ is an $M \times M$ linear transformation matrix. After the reservoir, the output state is read out through photon-number measurements, with the measurement outcomes represented as classical photon-count data. First- and second-order photon-number moments are then estimated. The photon-number operator for the $i$-th output mode is defined as $\hat{n}_i = \hat{b}_i^\dagger \hat{b}_i$, and the feature vector is

$$\boldsymbol{\phi}(\boldsymbol{x}) = (\langle \hat{n}_1 \rangle, \dots, \langle \hat{n}_M \rangle, \langle \hat{n}_1 \hat{n}_2 \rangle, \dots, \langle \hat{n}_k \hat{n}_l \rangle, \dots)^\mathsf{T}. \tag{2}$$

This measured feature vector is the classical output of the quantum reservoir and serves as the input to the linear readout. Its computational utility depends on two properties: nonlinear feature transformation and temporal memory. A nonlinear transformation maps the input data into a richer feature space, allowing a linear readout to represent nonlinear input–output relations[28]. Memory allows reservoir features to retain information about previous inputs, which is essential for temporal processing[29]. The following analysis identifies the physical origins of these two properties in quantum reservoir computing.

For the input state in Eq. (2), the first-order photon-number moment at output mode $k$ is

$$\langle \hat{n}_k \rangle = \sum_{i=1}^{M} |U_{ki}|^2 \, sinh^2 \, r_i \,, \tag{3}$$

where $U_{ki}$ is the matrix element in row $k$ and column $i$. For two distinct output modes $k$ and $l$ ($k \neq l$), the second-order photon-number moment is

$$\langle \hat{n}_k \hat{n}_l \rangle = \sum_{i=1}^{M} \sum_{j=1}^{M} |U_{ki}|^2 |U_{lj}|^2 \sinh^2 r_i \sinh^2 r_j + |\sum_{i=1}^{M} U_{ki} U_{li}^* \sinh^2 r_i|^2 + \frac{1}{4} |\sum_{i=1}^{M} U_{ki} U_{li} \sinh 2r_i|^2. \quad (4)$$

Substituting $r_i = x_i$ into Eqs. (3) and (4) make the nonlinear dependence of the measured photon-number moments on the input data explicit (see **Supplementary Note I**). The first-order moment inherits nonlinearity from the photon-number statistics of the squeezed input states. The second-order photon-number moment further combines responses from pairs of input modes. Thus, although the multimode interferometer transforms the optical field operators linearly, the encoding–interference–measurement process produces classical features that are nonlinear functions of the input.

Temporal memory is activated when the encoded data contains information from earlier cycles. Without feedback, past and present inputs are directly encoded in the squeezing parameters of the current input state, yielding memory determined by the supplied history window. With measurement-based feedback, selected photon-number features from the preceding cycle are re-encoded into the next input state, forming a recurrent loop that propagates information beyond the supplied window. In both cases, the measured feature vector carries a statistically accessible dependence on past inputs, as historical information is encoded into the quantum state. The passive unitary $U$ transforms this history-dependent state but does not itself store optical states between cycles.

Finally, a classical linear readout converts the measured features into the task output,

$$\hat{y} = \boldsymbol{W}\boldsymbol{\phi}(\boldsymbol{x}) + b, \quad (5)$$

where $\boldsymbol{W}$ and $b$ are learned classically. For static tasks, the readout serves as a classifier, while for dynamic tasks, it functions as a regressor predicting future values.

## Quantum reservoir computing microprocessor

Building on the theoretical framework, we fabricated a 16-mode quantum reservoir computing microprocessor on a silicon photonic platform. The device monolithically integrates an encoder comprising 16 single-mode squeezed-light sources with a reservoir core implemented by a fully programmable $16 \times 16$ linear-optical interferometer. This architecture physically realizes the sequence of squeezing-parameter encoding and multimode quantum-state transformation described above.

In the encoder, single-mode squeezed vacuum states are generated via degenerate spontaneous four-wave mixing in silicon spiral waveguides. Pump intensity controls the squeezing amplitude, thereby encoding each component of the classical input vector into the corresponding optical mode. Integrated asymmetric Mach–Zehnder interferometers (MZIs) suppress residual pump light and route the squeezed states into the reservoir core. The reservoir core is implemented using a Clements-mesh interferometer comprising 120 MZIs and 240 thermo-optic phase shifters. The splitting ratios and relative phases of the MZIs define the transfer matrix $U$, which mixes the independently encoded input modes into a multimode output state[30]. Once selected, the interferometer configuration remains unchanged during data acquisition and classical readout training. The output modes are coupled out of the chip and measured with an array of superconducting nanowire single-photon detectors. Time-to-digital conversion electronics record detection events as time-tag streams, from which single-mode photon counts and intermode coincidences are extracted. These experimental data yield the first- and second-order photon-number moments, forming the feature vector $\phi(x)$. The measured feature vector is then transferred to the CPU, where a trained linear readout generates the task output. Readout parameters are trained offline and subsequently loaded onto the CPU for inference. Details of the optical setup, device characterization, and squeezed-source performance are provided in **Supplementary Notes II-IV**.

The complete algorithm workflow is summarized in **Fig. 1c**: classical data are encoded into squeezing parameters, processed through the unaltered quantum circuit to generate a multimode Gaussian state, and converted into photon-statistical moments for classical linear readout. Together, these elements realize the full quantum reservoir computing pipeline, from quantum state encoding, reservoir computing core, and feature extraction, to classical linear inference, leveraging quantum dynamics to enhance feature extraction while retaining efficient classical trainability.

## Experimental demonstration of nonlinear mapping and memory

Reservoir computing requires a nonlinear mapping and fading memory for input data. The framework described above provides these two functions within a single photonic system. We first demonstrated nonlinear mapping and temporal memory separately, then evaluated their combined performance on a nonlinear temporal task.

The nonlinear mapping capability was assessed using a static two-dimensional XOR classification task from Eq. (6), a standard benchmark for nonlinear feature mapping. **Figure 2a** shows the dataset in the original input space, where samples from each class occupy diagonally opposite quadrants and cannot be separated by a single linear boundary. Following the experimental workflow, each input sample was encoded into the QPU without feedback, and features were extracted from photon-number measurements. These features were then fed into a linear readout implemented by logistic regression with a maximum of 500 iterations. As a control, the same readout was trained directly on the two raw input coordinates. **Figure 2b** shows that a linear classifier applied directly to the raw inputs achieved only 33.3% test accuracy. By contrast, when inputs were processed by the QPU, a linear readout of the measured photon statistics reached 98.3% accuracy (**Fig. 2c**). These results demonstrate that the quantum reservoir computer performs a nonlinear transformation, mapping the XOR data into a higher-dimensional statistical feature space in which the nonlinear relation becomes linearly accessible. Details of data generation, training, and accuracy evaluation are provided in **Methods** under "Static XOR data generation and evaluation".

Temporal memory was evaluated using an independent and identically distributed (IID) random input sequence, which removes correlations that could otherwise allow past inputs to be inferred without memory. The 16 input modes were divided into true-history modes (T), which encoded the current and recent inputs, and feedback modes (F), which received the previously measured reservoir output (**Fig. 2d**). Five allocations were examined, ranging from 1T+15F to 16T+0F. For each recall delay $d$, a separate ridge readout was trained to reconstruct the input presented *d* steps earlier from the 152 measured photon-counting features. Recall performance was quantified by the memory capacity $MC_d$ from Eq (11), defined as the squared Pearson correlation between the reconstructed and target sequences. Values near 1 indicate accurate recall, whereas values near 0 indicate little recoverable information. **Figure 2e** shows representative target and experimentally predicted values for the one-step memory task. Across 3 experimental repeats, the mean one-step memory capacity was highest for 4T+12F ($MC_1$ = 0.983±0.001), and lowest for 16T+0F ($MC_1$ = 0.608±0.0068). **Figure 2f** shows the memory capacity as a function of delay for the five T/F allocations. The 1T+15F configuration retained strong information over the first two delays, 4T+12F extended useful recall to four delays, 8T+8F to seven delays, and 12T+4F to elevan delays. The 16T+0F configuration maintained moderate recall across all fifteen tested delays. Further analysis of the respective contributions of true-history and feedback modes is provided in **Supplementary Note V**. These results show that the allocation of history and feedback modes tunes the balance between short-delay reconstruction accuracy and memory span: feedback-rich configurations favor accurate short-term recall, whereas history-rich configurations extend recall over longer windows. Experimental details are provided in **Methods** under "IID fading-memory task".

Finally, nonlinear mapping and temporal memory were evaluated jointly using a binary temporal XOR task from Eq. (13). At each time step, the target was defined as the XOR of the current and previous binary inputs. The task therefore required both memory of the previous input and nonlinear processing of consecutive inputs. Performance was evaluated on an independent chronological test sequence. For the 1T+15F allocation with feedback gain 0.15, the linear readout correctly classified 48 of 49 test samples, corresponding to 98.0% accuracy (**Fig. 2g**). The close agreement between target and prediction shows that feedback-based memory and nonlinear photonic feature mapping can operate together in a sequential task. Details of the input sequence and accuracy metric are provided in **Methods** under "Binary temporal XOR task".

These experiments demonstrate that the quantum reservoir computer architecture provides both the nonlinear mapping and temporal memory required for reservoir computing. The static XOR task confirms that photon-counting statistics yield nonlinear feature mappings that render linearly inseparable relations accessible to a linear readout. The IID delayed-recall task shows that history/feedback allocation tunes both memory strength and span. The temporal XOR task demonstrates that these functions can be combined to process nonlinear sequential relations. The same photonic reservoir therefore provides nonlinear mapping and programmable fading memory while requiring training only at the linear readout.

## Experimental demonstration of versatile tasks

After characterizing nonlinear mapping and temporal memory, we evaluated the versatility of the experimental quantum reservoir computer using two representative learning tasks: multiclass classification with the Iris dataset and prediction of Mackey–Glass dynamics. These tasks probe static inference and continuous temporal processing, respectively.

Static classification performance was assessed using the widely adopted Iris dataset, a standard benchmark for multi-class learning. Each Iris sample consists of 4 real-valued features, and the dataset was split into training and test sets at 70%/30% (see **Methods**, "Iris classification"). **Figure 3a** shows the confusion matrix on the test set. A logistic regression model trained on the measured photon statistics correctly classified 42 of 45 samples, yielding an accuracy of 93.33%. Predictions were concentrated along the diagonal, with only a few off-diagonal entries, confirming that the three classes are well separated in the reservoir-generated feature space. Optimization traces show that training and test accuracies stabilized after approximately 50 L-BFGS steps (**Fig. 3b**), with corresponding log losses reaching stable values (**Fig. 3c**). These results confirm that the final classification accuracy reflects convergence of the linear readout.

Temporal prediction capability was evaluated using the Mackey–Glass time series, generated by a nonlinear delay differential equation and widely used as a benchmark for forecasting nonlinear dynamical systems (see **Methods**, "Mackey–Glass task"). Its evolution depends on both the current state and a delayed past state., with the delay parameter $\tau$ determines how far into the past this dependence extends and changes the temporal correlations and dynamical behaviour of the generated sequence. Prediction performance was quantified by the mean squared error (MSE) from Eq (16) between predicted and target values. For $\tau = 17$, the experimental prediction closely tracked the target over the chronological test set (**Fig. 3d**), with predicted and target values concentrated around the identity line and an MSE of 0.0101 (**Fig. 3e**). The reconstructed attractor retained the characteristic phase-space geometry of the target dynamics (**Fig. 3f**). Robustness was further examined by varying $\tau$ from 20 to 50. The interferometric reservoir remained unchanged, while separate linear readouts were trained for each sequence. Across the tested regimes, the MSE remained between 0.0056 and 0.0124 (**Fig. 3g**). These results show that the photonic reservoir maintains consistently low one-step prediction errors across Mackey–Glass regimes with different degrees of chaotic complexity, showing that its temporal processing capability is not confined to a single dynamical regime.

Prediction stability is another important measure of system performance, which was assessed using a free-running test under closed-loop operation[31]. Unlike supervised prediction, where true history is always supplied, free-running forecasts recursively feed previous predictions back as inputs, amplifying errors and eventually driving trajectories away from the true dynamics. This stringent test evaluates whether the reservoir has genuinely captured the dynamical structure. Intermediate prediction outputs were recursively fed back as subsequent inputs, i.e., $\hat{x}_{t+\Delta} \rightarrow x_{t+\Delta}^{(\mathrm{in})}$, where $x_{t+\Delta}^{(\mathrm{in})}$ denotes the input used at the next prediction step. **Figure 4a** shows a 200-step free-running prediction, with cumulative MSE presented in **Fig. 4b**. During the initial stage, the predicted trajectory tracked the true sequence well, with cumulative error below 0.2 over the first 27 steps. Thereafter, error gradually accumulated and the predicted trajectory deviated from the true evolution. Although error growth is unavoidable in long-horizon autonomous rollouts, this result demonstrates that the platform is not limited to short-term fitting under teacher forcing, but can support autonomous prediction over finite horizons while preserving effective reconstruction of the target dynamical structure.

## Computing with a zero-power reservoir core

For integrated photonic platforms, whether the classification and prediction capabilities of the quantum photonic reservoir depend on active tuning of the on-chip interferometric circuit has direct physical and engineering relevance. Reservoirs requiring active phase tuning (non-zero bias states) face rapid increases in system complexity, control power consumption, and thermal-management overhead as they scale[32]. By contrast, if a reservoir operating without active phase tuning (zero-bias state) can already achieve high performance, then the control complexity and overhead of the reservoir core can be significantly reduced. Motivated by this consideration, we compare reservoir performance in the zero-bias state, where all on-chip MZI phase shifters are driven at zero current, with random-bias states with non-zero driving currents. Random-bias configurations were obtained by sampling random unitary matrices with different seeds and mapping them to phase parameters of the MZI circuit via Takagi decomposition[33]. Apart from the reservoir, all other aspects of the experiment were kept unchanged, including input encodin, feature extraction, and readout training. This comparison therefore directly tests whether active control is necessary for useful computation under identical task and readout conditions.

The power consumption of the reservoir core is estimated as $P_{\mathrm{RC}} = \sum_i I_i^2 R_i$, where $I_i$ and $R_i$ are the driving current and resistance of the $i$-th thermo-optic phase shifter, respectively. This quantity characterizes the power required to maintain the reservoir configuration, but not the total system power. In the zero-bias state, all phase shifters were driven at zero current, yielding $P_{\mathrm{RC}} = 0$. By contrast, any non-zero-bias state incurred additional control-power overhead. **Figure 5** compares the performance of the reservoir core in the zero-bias and non-zero-bias states. **Figure 5a** shows Iris classification accuracy as a function of reservoir-core power. Although accuracy varied across configurations, the zero-power state maintained competitive performance. **Figure 5b** shows test-set MSE in the Mackey–Glass prediction task versus MZI power consumption. The zero-bias state achieved an MSE of approximately 0.011, lower than or comparable to non-zero-bias configurations (0.012–0.018). One possible explanation for the improved performance is suppression of thermal crosstalk across the photonic chip, leading to enhanced system stability. These results indicate that the zero-bias regime not only enables operation at zero current, but also eliminates active power consumption while preserving strong computational capability.

We show that comparable performance with and without active tuning arises from the unique encoding of classical data into multimode Gaussian states, fully characterized by their covariance matrices (see **Supplementary Note VI**). Classical inputs encoded in squeezing parameters are mapped to a specific Hilbert-space sector via a random unitary transformation. Subsequent random unitaries generally map this sector to another, while preserving global features of the encoded data. This mechanism guarantees comparable classification and prediction performances. Under a uniform-loss assumption, the covariance matrix is simply rescaled, leaving encoded information unaffected. This further implies intrinsic robustness of the quantum reservoir computing architecture against photon loss.

Both theoretical analysis and experimental data demonstrate that the quantum photonic reservoir operates effectively without fine tuning. In the zero-bias state, the fixed interferometric structure already provides an effective mapping from inputs to features, supporting both static classification and dynamic prediction. Thus, computational capability emerges from the interplay of squeezed-state encoding, multimode interferometric mixing, and statistical readout, rather than from continuously adjusted parameters. This result carries a clear implementation-level implication: active programming overhead of the reservoir core can be reduced to zero while preserving strong task performance. For integrated photonic platforms scaling to larger mode numbers, longer operating times or more demanding tasks, diminished reliance on active tuning offers an immediate and practical advantage.

## Conclusion and outlook

This work experimentally demonstrates the first integrated photonic quantum reservoir computer that unites nonlinear feature mapping, tunable temporal memory, task-flexible processing, and ultra-low active-control overhead. Static XOR and fading-memory experiments establish the presence of nonlinearity and memory, while temporal XOR task verifies their joint operation. Iris classification and Mackey–Glass prediction further showcase the versatility of architecture across both static and dynamical learning tasks.

Crucially, the passive chip retains classification and prediction capabilities without active biasing, eliminating the need for continuous internal phase tuning. This property addresses a central challenge in photonic machine learning: achieving nonlinearity, memory, and task flexibility without incurring substantial increases in hardware complexity, control overhead, or power consumption. The principle opens a route to repurposing large-scale photonic quantum processors as reservoirs without reconfiguring their internal optical networks, thereby accessing additional modes and intermode correlations to expand the computational feature space. It also changes design requirements for photonic integrated chips: a passive reservoir core could eliminate the need for individually addressed heaters and associated wiring, reducing static power dissipation and thermal crosstalk. This would allow optical networks to scale in size without electrical control of every internal mixing element.

The physical mechanism underlying this architecture arises from the combined action of quantum-state encoding, multimode interferometric mixing, and photon statistical readout, which together map classical data onto photon correlations. Although the interferometric circuit itself is strictly linear, the complete transformation from classical inputs to reservoir features exhibits an effective nonlinearity. This distinguishes the present architecture from many existing optical-neural-network implementations, where nonlinearity must be supplied by Kerr elements, activation units, or repeated optoelectronic conversion. In this sense, the present platform points to a quantum-enabled advantage rooted in the physical mechanism itself: quantum-state encoding and statistical measurement provide a natural route to task-useful nonlinearity and high-dimensional feature generation.

The demonstrated nonlinear mapping, tunable temporal memory, dynamic prediction capability, absence of active tuning, and low computational power consumption together satisfy the core requirements of quantum reservoir computing for signal classification, time-series prediction, dynamical control, and surrogate modelling of complex systems. These capabilities further point toward broader applications, including optical communications, next-generation optical networks, edge inference for the Internet of Things, green data centres, intelligent robotics, and AI for science and digital twins. Overall, this work establishes that the combination of quantum-state encoding, an on-chip multimode interferometric circuit, and photon-statistical readout provides a scalable quantum reservoir computing architecture with a clear physical basis, ultra-low power consumption, and strong potential for large-scale quantum computation.

## Method

### Static XOR data generation and evaluation

The static XOR task was used to evaluate nonlinear feature mapping. Each input sample $i$ contained two real-valued coordinates, $X_i = (x_{i,1}, x_{i,2})$. The class label was assigned according to

$$y = \begin{cases} 1, & x_{i,1}x_{i,2} > 0, \\ 0, & x_{i,1}x_{i,2} < 0. \end{cases} \tag{6}$$

Samples lying on either coordinate axis were excluded. Consequently, samples in diagonally opposite quadrants belonged to the same class, whereas adjacent quadrants belonged to different classes.

The two coordinates were encoded into two optical input modes. The remaining modes were set to 0. Each sample was therefore processed independently. Photon-counting measurements were converted into the first- and second-order statistical features defined above. A linear logistic regression model was then trained on the measured quantum reservoir computing features. As a control, the same readout was trained directly on the raw input coordinates using identical training and test partitions.

Classification performance was evaluated on an independent test set. The test accuracy was calculated as

$$Accuracy = \frac{N_{correct}}{N_{test}}, \tag{7}$$

where $N_{\mathrm{correct}}$ is the number of correctly classified test samples and $N_{\mathrm{test}}$ is the total number of test samples.

### IID fading-memory task

Temporal memory was quantified using an IID sequence

$$u_k \sim \mathcal{U}(-1, 1). \tag{8}$$

Statistical independence between successive inputs ensured that past values could not be inferred from correlations within the input sequence. The reconstruction of $u_{k-d}$ therefore provided a direct measure of the memory retained by the quantum reservoir computing.

The 16 optical modes in the encoder were divided into T history modes and F feedback modes, with T + F = 16. At time step k, the encoded input vector was

$$m_k = (u_k, u_{k-1}, \dots, u_{k-T+1}, g\phi_{k-1,1}, \dots, g\phi_{k-1,F}), \quad (9)$$

where the history modes contained the current input and the preceding T-1 inputs. The feedback modes received the normalized first-order photon statistics obtained at the immediately preceding measurement. The feedback gain was g=0.15. Missing history values at the beginning of a trajectory were set to zero. All encoded coordinates were restricted to [-1,1]. Five mode allocations were evaluated, namely 1T+15F, 4T+12F, 8T+8F, 12T+4F and 16T+0F. The interferometric reservoir configuration was kept fixed for all allocations

Each trajectory contained 285 time steps. The first 30 measurements were discarded as washout. A further 15 samples were reserved for delay alignment. The next 160 samples formed the training set, and the final 80 samples formed a chronological test set. Delayed reconstruction was evaluated for d=1, … ,15, with the target

$$y_k{}^{(d)} = u_{k-d}. \quad (10)$$

A separate ridge readout was trained for each recall delay and mode allocation. The measured features were normalized within each measurement and standardized using statistics obtained from the training set.

The memory capacity at delay d, denoted by $MC_d$, was calculated as the squared Pearson correlation between the target and reconstructed sequences, expressed as

$$MC_d = corr^2\left(y^{(d)}, \hat{y}^{(d)}\right) \quad (11)$$

where $y^{(d)}, \hat{y}^{(d)}$ denote the target and reconstructed sequences at delay d, respectively. The total memory over the measured delay range was defined as

$$MC_{total} = \sum_{d=1}^{15} MC_d \quad (12)$$

**Binary temporal XOR task**

The temporal XOR task used a random binary input sequence $s_k \in \{0,1\}$, with 0 and 1

occurring with equal probability. The target output at time step k was defined as

$$y_k = s_k \oplus s_k, \tag{13}$$

where $\oplus$ denotes the XOR operation. The target depends jointly on the present and preceding inputs. This task required both temporal memory and a nonlinear transformation.

The experiment used the 1T+15F allocation with a feedback gain of g = 0.15. The single history mode encoded the current input, while the remaining 15 modes received normalized first-order photon statistics from the preceding measurement. A sequence of 158 samples was measured: 10 discarded as washout, 99 used for training, and 49 for testing. Features were normalized per measurement and standardized using training parameters. An unregularized linear regression model was used as the primary readout. Accuracy was evaluated according to the definition provided for the static XOR task.

**Iris classification**

Static multiclass classification was evaluated using the Iris dataset. The dataset contains 150 samples from three plant species, with 50 samples per class. Each sample is described by sepal length, sepal width, petal length and petal width. The four input features were normalized to [-1,1] using the procedure described above and encoded into four optical modes. The remaining modes were set to 0. Each sample was therefore processed independently.

The measured photon-statistical features were used to train a multinomial logistic-regression readout with a maximum of 500 iterations. The dataset was divided into 105 training samples and 45 test samples, corresponding to a 70%/30% split. Performance was evaluated using the classification accuracy defined for the static XOR task.

**Mackey–Glass task**

Nonlinear time-series prediction was evaluated using the Mackey–Glass delay differential equation, expressed as

$$\frac{dx(t)}{dt} = \frac{\beta x(t - \tau)}{1 + x^n(t - \tau)} - \gamma x(t), \tag{14}$$

where $\beta = 0.2$, $\gamma = 0.1$ and $n = 10$ and the output of $x(t)$ is normalized to the interval [-1,1]. The delay parameter $\tau$ controls the temporal complexity of the generated dynamics. The equation was integrated by the forward Euler method using a step size of $dt = 0.1$. At each integration step, the next value was calculated as

$$x_{j+1} = x_j + \left[\frac{\beta x_{j-\tau_{step}}}{1 + x_{j-\tau_{step}}^n} - \gamma x_j\right] dt, \tag{15}$$

where $\tau_{step} = round\left(\frac{\tau}{dt}\right)$.

A ridge readout was trained on the chronological training segment and evaluated on the subsequent test segment. The ridge coefficient used for the reported experiment was $\alpha = 10^{-6}$. Mean squared error (MSE) serves as the metric for assessing prediction performance, which is defined as

$$\text{MSE} = \frac{1}{N}\sum_{t=1}^{N}(\hat{y}_{t+\Delta} - y_{t+\Delta})^2 . \tag{16}$$

Where $y_{t+\Delta}$ is the target value, $\hat{y}_{t+\Delta}$ is the corresponding experimental prediction, $N$ is the number of test samples and $\Delta$ is the prediction horizon. One-step prediction corresponds to $\Delta = 1$.

**Acknowledgments:** This research was conducted at the JC STEM Lab of Quantum Technology, funded by The Hong Kong Jockey Club Charities Trust. This work was also supported by Hong Kong Polytechnic University (1-CD3J, 1-BEAZ, 1-BDWT, 1-BEDH, 1-BEET) and by Hong Kong Research Grant Council/University Grants Committee (grant no. 21203724).

**Author contributions:** W.W. and A-Q.L. conceived the project. W.W. and Z.T. performed the experiments. W.W. and D.Q.S. carried out the simulations and performed the theoretical analysis. W.W. and A-Q.L. managed the project. W.W., Z.T., D.Q.S., L.K.C., H.C., M.L.F., M.G., J.T, H.C, L-C.K, and A-Q.L. discussed and wrote the manuscript with input from all authors. All authors discussed the results and contributed to the manuscript.

**Competing interests:** The authors declare no competing interests.

**Data availability:** The data supporting this study's findings are available from the corresponding authors upon reasonable request.

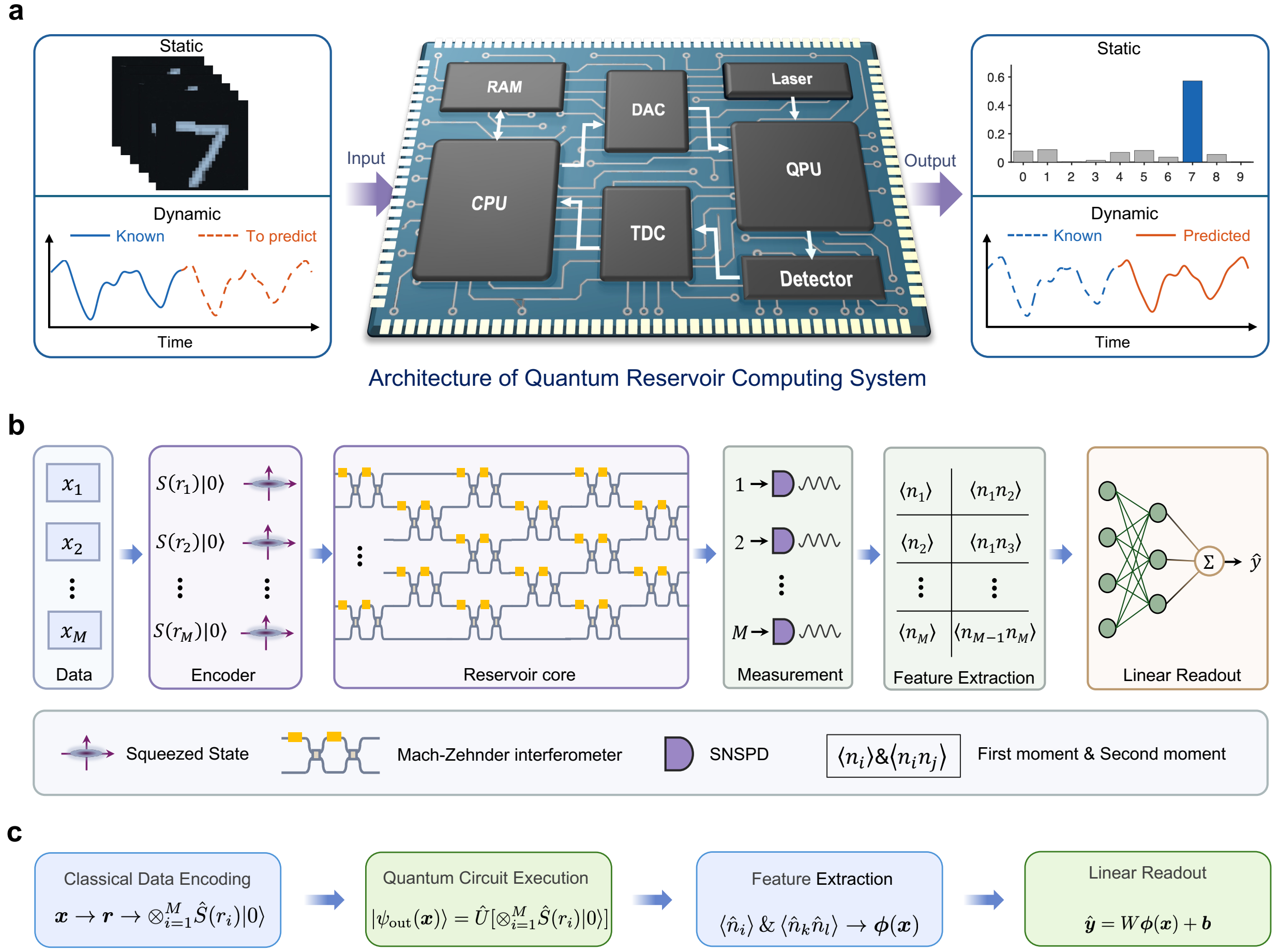

**Fig. 1 Quantum reservoir computer based on a photonic quantum microprocessor (QPU). a,** Conceptual system-level architecture showing task data input, quantum-state encoding and reservoir processing in the QPU, and classical linear readout. Representative tasks include static classification (top: handwritten digit "7" mapped to a probability distribution peaking at 7) and dynamic time-series prediction (bottom: known history in blue, predicted trajectory in orange closely matching the ground-truth future segment). **b,** Detailed workflow of the photonic QPU pipeline, including input encoding into squeezed states, multimode interferometric mixing in the reservoir core, and photon-statistical measurement for feature extraction. **c,** Full algorithmic workflow summarizing the sequence of classical data encoding, quantum circuit execution, feature extraction, and linear readout.

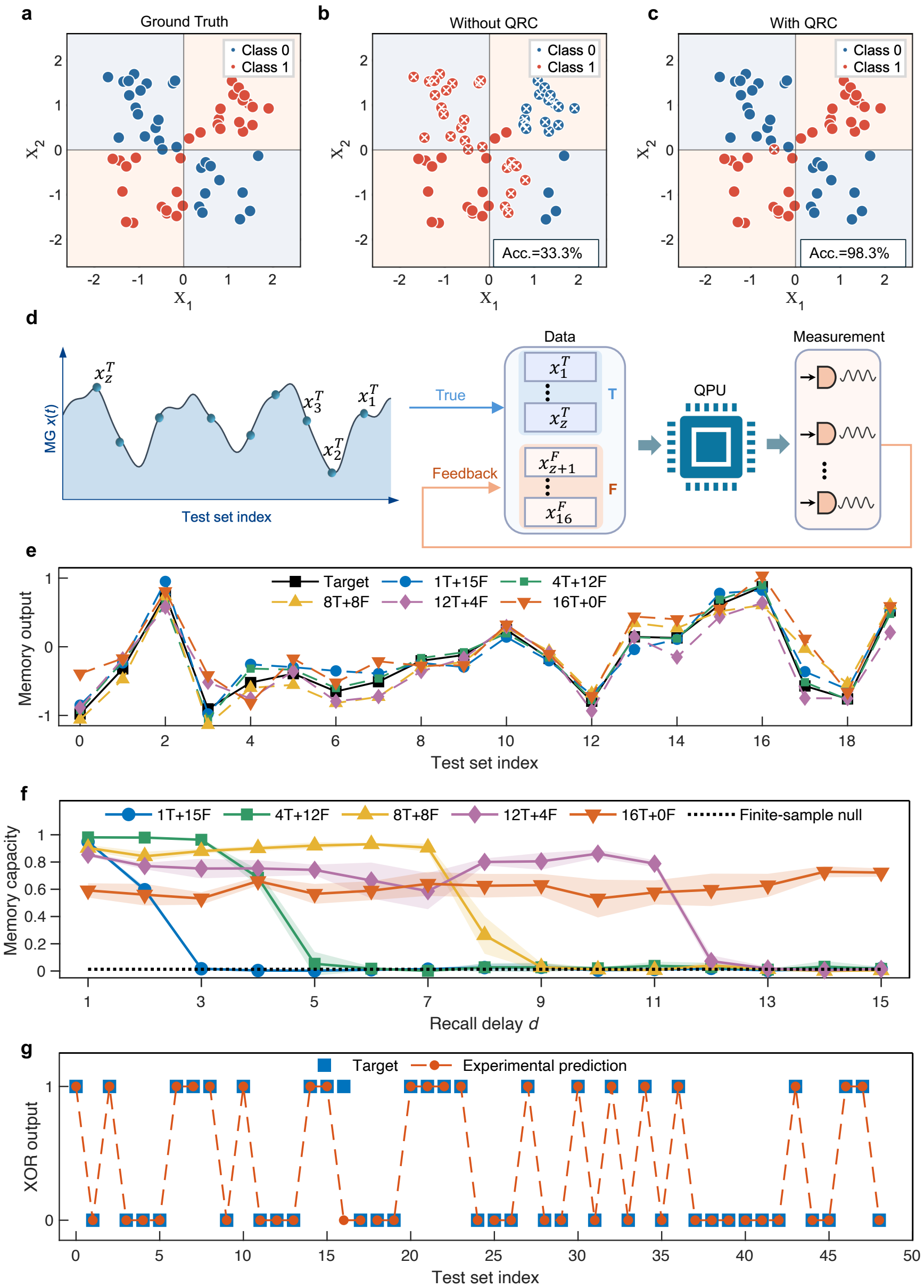

**Fig. 2 Nonlinear mapping and fading memory in the QPU. a,** Ground truth labels for the static XOR dataset. **b,** Predictions from a linear classifier trained directly on the raw input coordinates, yielding 33.3% test accuracy. **c,** Predictions from a linear readout trained on photon-statistical features extracted by the QPU, achieving 98.3% test accuracy. **d,** Schematic of history and feedback mode allocation. Sixteen input modes are divided into T history modes encoding past inputs and F = 16 – T feedback modes receiving measurement outputs from the preceding cycle. **e,** Representative one-step delayed reconstructions showing target and predicted outputs for five different T/F allocations. **f,** Memory capacity versus recall delay, quantified by squared Pearson correlation between target and prediction. Solid lines show the mean across six acquisitions; shaded regions indicate one standard deviation; dotted line marks the finite-sample baseline. **g,** Binary temporal XOR task results. Targets and experimental predictions for the chronological test set are shown. An unregularized linear readout with threshold 0.5 correctly classified 48 of 49 test samples.

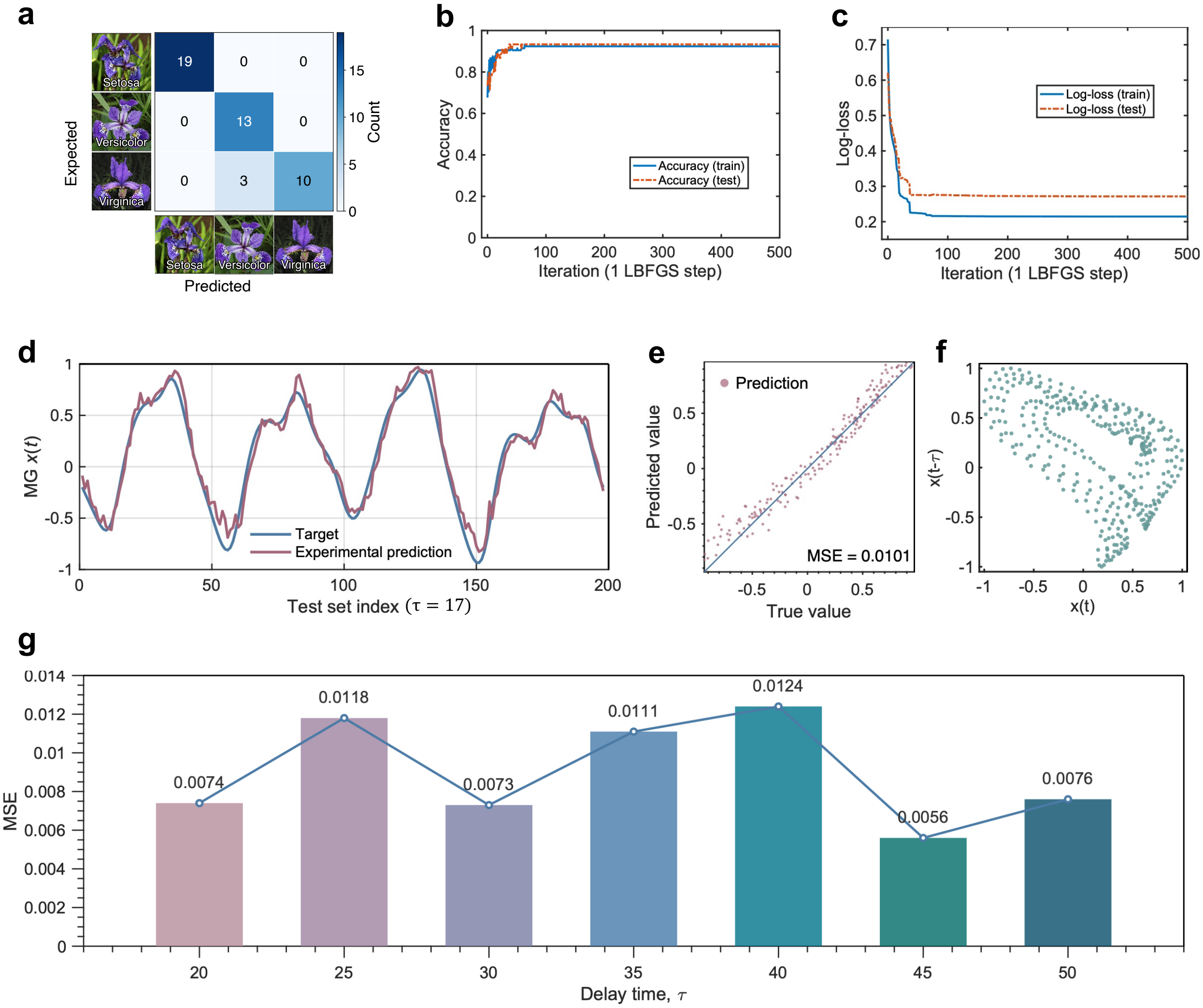

**Fig. 3 Versatile task performance of the quantum reservoir computer. a,** Confusion matrix for Iris dataset classification using photon-statistical features generated by the reservoir, yielding 93.3% test accuracy (42 of 45 samples correctly classified). **b, c,** Training and test accuracy and log loss as functions of L-BFGS optimization steps, showing convergence after ~50 iterations. **d,** Representative one-step prediction of a Mackey–Glass time series with delay parameter $\tau$ = 17; target sequence in blue and experimental prediction in red. **e,** Predicted versus target values for the test sequence in d; points cluster along the diagonal, with mean squared error (MSE) of 0.0101. **f,** Delayed-coordinate attractor reconstructed from the experimentally predicted Mackey–Glass sequence at $\tau = 17$, retaining the characteristic phase-space geometry. **g,** One-step prediction errors for Mackey–Glass sequences with delay parameters $\tau = 20$ to 50. MSE remains between 0.0056 and 0.0124 across regimes, demonstrating robust temporal processing across different dynamical complexities.

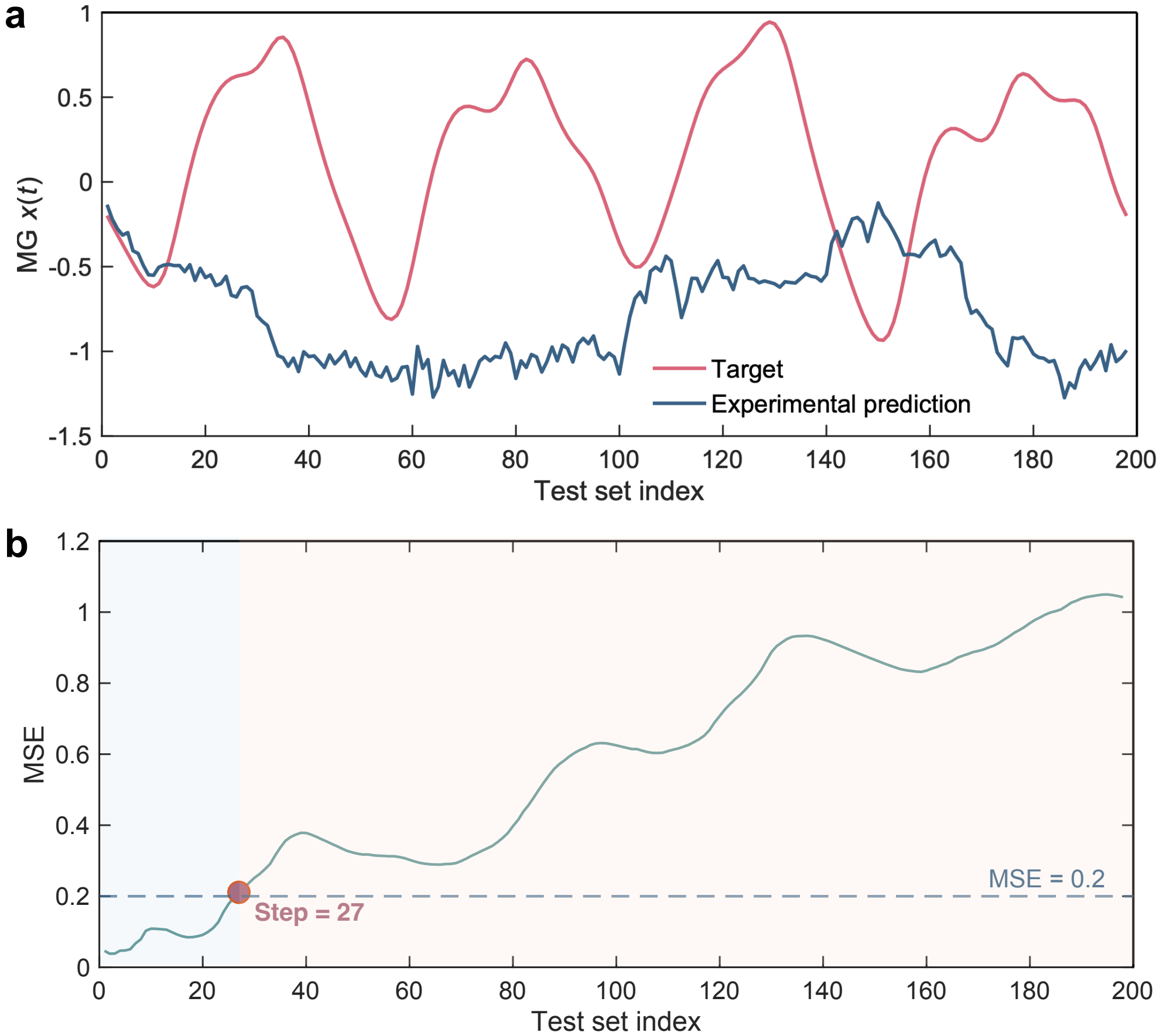

**Fig. 4 Free-running prediction under closed-loop operation. a,** Target Mackey–Glass sequence (red) and corresponding 200-step experimental free-running prediction (blue). After initialization, each intermediate prediction is recursively fed back as input to the next step, without access to future target values. **b,** Cumulative MSE between prediction and target sequence. The dashed line marks the 0.2 error threshold used to define the prediction horizon. The cumulative error remains below this threshold for the first 27 steps, after which deviations grow. The red marker indicates the threshold crossing.

**a**

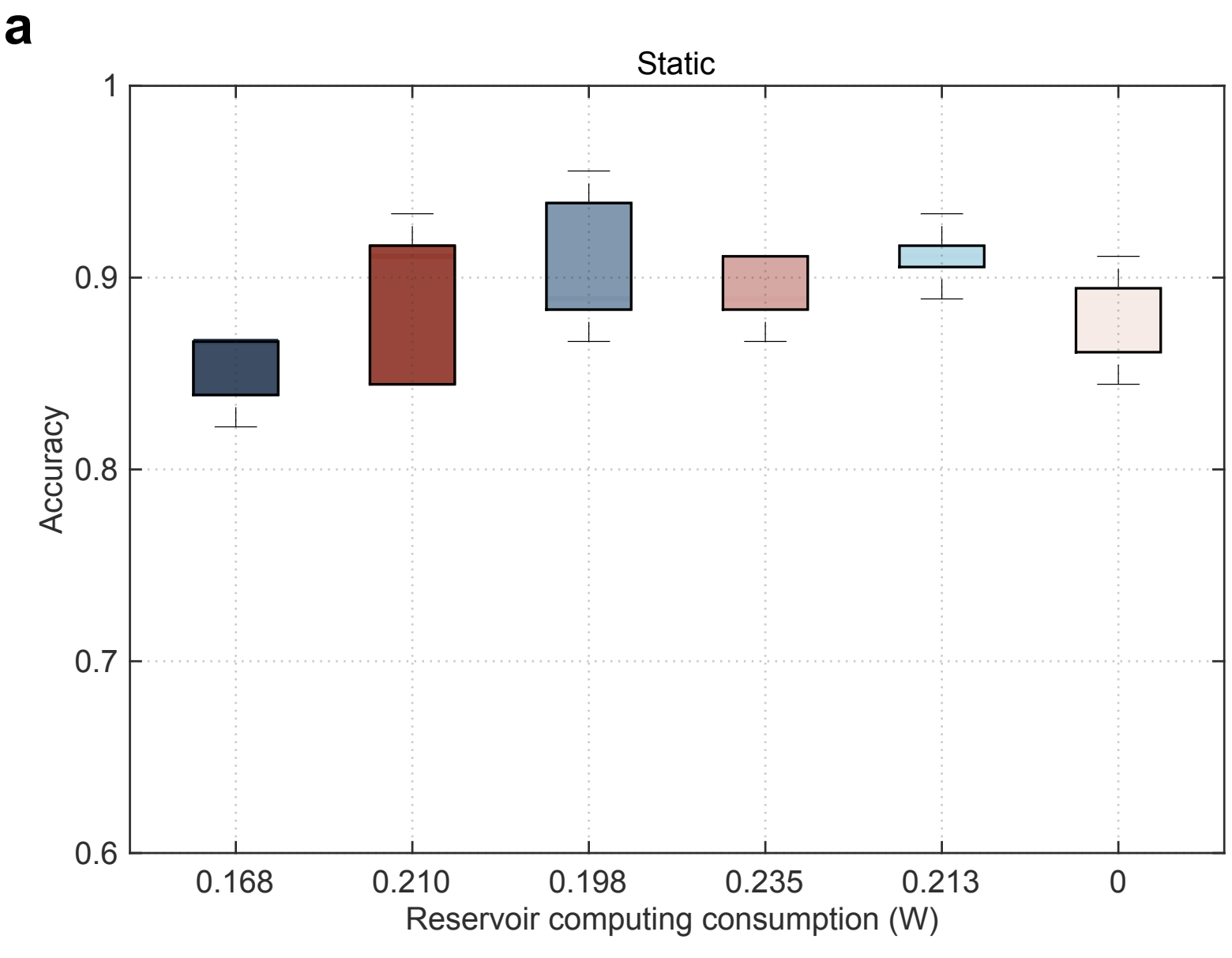

**b**

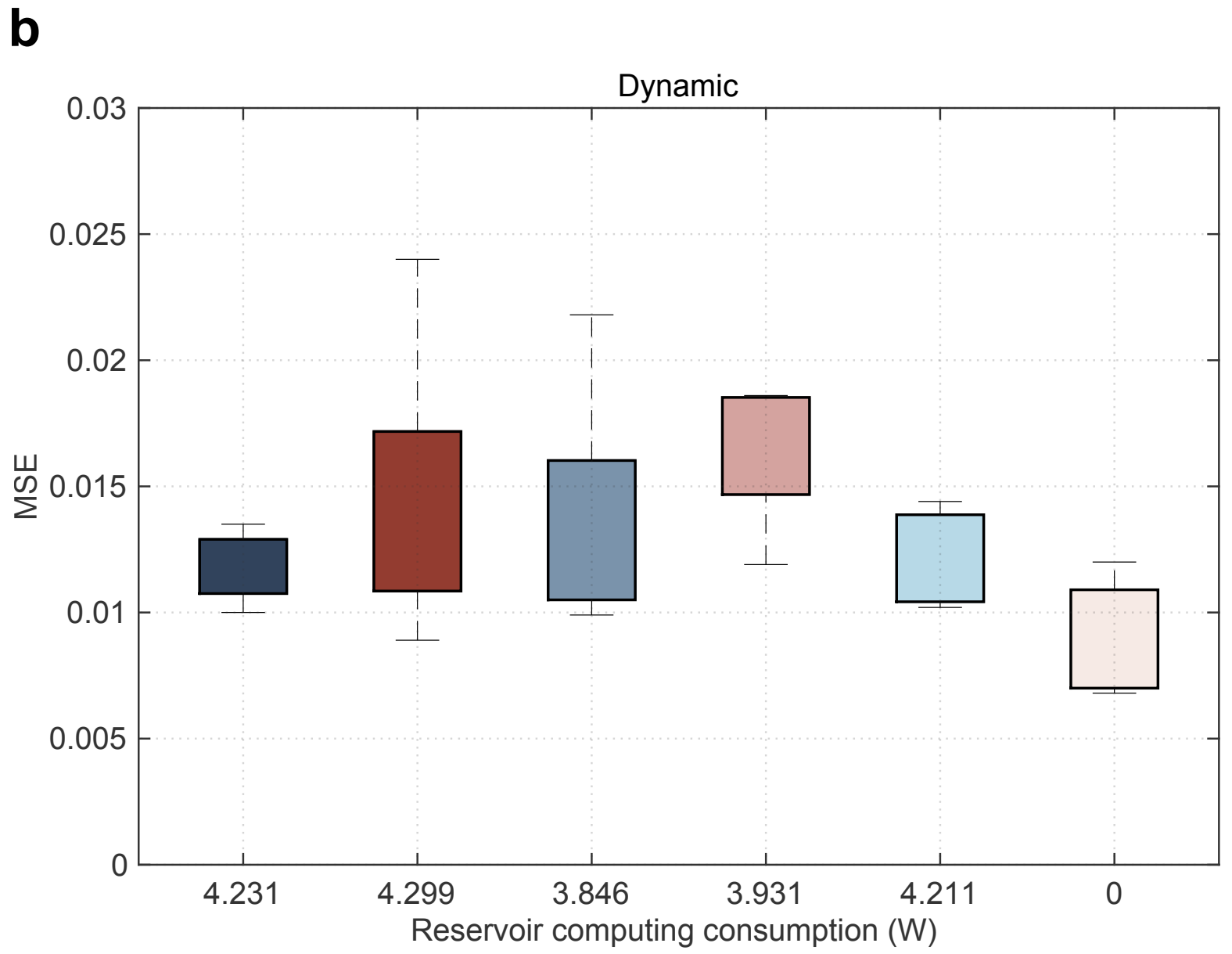

**Fig. 5 Quantum reservoir computer performance under zero-bias and randomly biased configurations. a,** Test accuracy for static Iris classification as a function of active phase-control power dissipated by the interferometric reservoir core. **b,** Test MSE for Mackey–Glass time-series prediction under the same comparison. The five non-zero-power operating points correspond to randomly biased interferometric configurations. The zero-power point denotes the zero-bias state, in which all on-chip phase shifters are unpowered. For each configuration, the center line indicates the median across five experimental trials, the boxes represent the interquartile range, and the whiskers show the minimum and maximum values.